\documentclass[aps,prb,singlecolumn,superscriptaddress,ecnumarabic,amssymb,
nobibnotes,preprint]{revtex4-2}

\usepackage[T1]{fontenc}     % Ensures proper font encoding
\usepackage{siunitx}
\DeclareSIUnit{\bar}{bar}
\DeclareSIUnit{\angstrom}{\text{Å}}

\usepackage{graphicx,xcolor}

\usepackage{xspace}
\usepackage{hyperref}
\usepackage{amsmath}
\newcommand{\bvec}[1]{\boldsymbol{#1}}

\newcommand{\MT}[1]
{\color{red} #1 \color{black}}
\newcommand{\NbSe}{NbSe\textsubscript{2}\xspace}

\newcommand{\dIdV}{$\mathrm{d}I/\mathrm{d}V$\xspace}

\newcommand{\Devi}{$2\times\!2$\xspace}
\newcommand{\Devii}{$\sqrt{3}\times\!\sqrt{3}\ \mathrm{R}30^{\circ}$\xspace}
\newcommand{\SoD}{$\sqrt{13}\times\!\sqrt{13}\ \mathrm{R}13.9^{\circ}$\xspace}
\newcommand{\C}{$C_3$\xspace}
\newcommand{\Supple}{Suppl.\ Mat.~}

\begin{document}

%\title{Superlattice engineering in van der Waals heterostructures}

%\title{Symmetry engineering in  van der Waals superlattices}
\title{Tailoring spontaneous symmetry breaking in engineered van der Waals superlattices} % Change the title back

%\title{Symmetry breaking in engineered van der Waals superlattices} % Change the title back

\author{Keda Jin}
\email{k.jin@fz-juelich.de}
\affiliation{Peter Gr\"unberg Institut (PGI-3), Forschungszentrum J\"ulich, 52425 J\"ulich, Germany}
\affiliation{J\"ulich Aachen Research Alliance, Fundamentals of Future Information Technology, 52425 J\"ulich, Germany}
\affiliation{Institut f\"ur Experimentalphysik II B, RWTH Aachen, 52074 Aachen, Germany}

\author{Lennart Klebl}
\affiliation{Institut für Theoretische Physik und Astrophysik and Würzburg-Dresden Cluster of Excellence ctd.qmat, Universität Würzburg, 97074 Würzburg, Germany}
%\affiliation{I. Institute for Theoretical Physics, Universität Hamburg, Notkestraße 9-11, 22607 Hamburg, Germany}

\author{Zachary A. H. Goodwin}
\affiliation{Department of Materials, University of Oxford, Parks Road, Oxford OX1 3PH, United Kingdom}
\affiliation{John A Paulson School of Engineering and Applied Sciences, Harvard University, Cambridge, MA, USA}

\author{Junting Zhao}
\affiliation{Peter Gr\"unberg Institut (PGI-3), Forschungszentrum J\"ulich, 52425 J\"ulich, Germany}
\affiliation{J\"ulich Aachen Research Alliance, Fundamentals of Future Information Technology, 52425 J\"ulich, Germany}
\affiliation{Institut f\"ur Experimentalphysik II B, RWTH Aachen, 52074 Aachen, Germany}

%\author{Tobias Wichmann}
%\affiliation{Peter Gr\"unberg Institut (PGI-3), Forschungszentrum J\"ulich, 52425 J\"ulich, Germany}
%\affiliation{J\"ulich Aachen Research Alliance, Fundamentals of Future Information Technology, 52425 J\"ulich, Germany}
%\affiliation{Institut f\"ur Experimentalphysik IV A, RWTH Aachen, 52074 Aachen, Germany}

%\author{(Stefan F. Tautz)}
%\affiliation{Peter Gr\"unberg Institut (PGI-3), Forschungszentrum J\"ulich, 52425 J\"ulich, Germany}
%\affiliation{J\"ulich Aachen Research Alliance, Fundamentals of Future Information Technology, 52425 J\"ulich, Germany}
%\affiliation{Institut f\"ur Experimentalphysik IV A, RWTH Aachen, 52074 Aachen, Germany}

\author{Felix L\"upke}
\affiliation{Peter Gr\"unberg Institut (PGI-3), Forschungszentrum J\"ulich, 52425 J\"ulich, Germany}
\affiliation{J\"ulich Aachen Research Alliance, Fundamentals of Future Information Technology, 52425 J\"ulich, Germany}
\affiliation{Institute of Physics II, Universität zu Köln,
Zülpicher Straße 77, 50937 Köln, Germany}

\author{Dante M. Kennes}
\affiliation{Institut  f\"ur Theorie der statistischen Physik, RWTH Aachen, 52074 Aachen}
\affiliation{J\"ulich Aachen Research Alliance, Fundamentals of Future Information Technology, 52425 J\"ulich, Germany}
\affiliation{Max Planck Institute for the Structure and Dynamics of Matter, Center for Free Electron Laser Science, 22761 Hamburg, Germany}

\author{Jose Martinez-Castro}
%\email{j.martinez@fz-juelich.de}
\affiliation{Peter Gr\"unberg Institut (PGI-3), Forschungszentrum J\"ulich, 52425 J\"ulich, Germany}
\affiliation{J\"ulich Aachen Research Alliance, Fundamentals of Future Information Technology, 52425 J\"ulich, Germany}
\affiliation{Institut f\"ur Experimentalphysik II B, RWTH Aachen, 52074 Aachen, Germany}

\author{Markus Ternes}
\email{m.ternes@fz-juelich.de}
\affiliation{Peter Gr\"unberg Institut (PGI-3), Forschungszentrum J\"ulich, 52425 J\"ulich, Germany}
\affiliation{J\"ulich Aachen Research Alliance, Fundamentals of Future Information Technology, 52425 J\"ulich, Germany}
\affiliation{Institut f\"ur Experimentalphysik II B, RWTH Aachen, 52074 Aachen, Germany}

%===============================================================================
\begin{abstract}
%===============================================================================
Superlattice engineering in van der Waals heterostructures (e.\,g.\ by moir\'e engineering) provides a powerful platform for designing electronic bands and realising correlated and topological quantum phenomena. 
%While twist-angle-induced moiré patterns are widely employed, 
Here, we pioneer a scheme to tailor superpotentials based on intrinsic substrate electronic orders. 
We show that this establishes a robust, self-aligned, and highly versatile route to band-structure control as we demonstrate in graphene by engineering two distinct, nearly commensurate superlattices using the charge density waves of 1T-\NbSe.
In these superlattices the graphene’s Dirac cones are folded either to the $\Gamma$-point or to the K-points of the mini-Brillouin zone. Using scanning tunnelling microscopy, we observe that the $\Gamma$-folded system preserves \C symmetry, while the K-folded system exhibits spontaneous symmetry breaking. Combining density functional theory with an interlayer interaction model, we reveal that this difference is not electronically driven but originates from a structural instability. Our work establishes superlattice engineering for designer quantum states and unveils a structural mechanism for controlled emergent symmetry breaking.

%Superlattice engineering in van der Waals heterostructures (vdW) provides a powerful platform for designing electronic bands and realising correlated and topological quantum phenomena. While twist-angle-induced moiré patterns are widely employed, proximity-induced superlattices based on intrinsic electronic orders, such as charge density waves (CDWs), offer a complementary route. Here, we demonstrate control over band-folding pathways in graphene by engineering two distinct, nearly commensurate superlattices using the CDW of 1T-\NbSe, which fold graphene’s Dirac cones either to the $\Gamma$-point or to the K-points of the mini-Brillouin zone (mBZ).
%, which results to selecting different symmetry outcomes. 
%(MT: This is already saidin the next sentence!)
%Using scanning tunnelling microscopy, we observe that the $\Gamma$-folded system preserves \C symmetry, while the K-folded system exhibits spontaneous symmetry breaking. Combining density functional theory with an interlayer interaction model, we reveal that this difference is not electronically driven but originates from a structural instability. Our work establishes a strategy for engineering quantum states via superlattice engineering and unveils a structural mechanism for emergent symmetry breaking.

% \KJ{Nature material requires the abstract to have max. 150 words, but the current version has 200 words.}
\end{abstract}

\maketitle
%===============================================================================
\section*{Main}
%===============================================================================
%\KJ{For the article: (1) Main text – up to 3,000 words, excluding abstract, Methods, references and figure legends. (2) The main section should separate to results and discussion.}
%\subsection{Introduction}
% A superlattice is a periodic modulation of a material property on a length scale larger than its underlying crystal lattice, which can reshape a material's electronic structure~\cite{Esaki1970,Park2008Anisotropic}. 

Superlattices in van der Waals (vdW) heterostructures  offer a way to reshape the electronic structure of two-dimensional (2D) materials and stabilise interaction-driven phases that are not present in the individual constituents~\cite{ Bistritzer2011moire, ponomarenko2013cloning-b61, Cao2018correlated, checkelsky2024flat-c0b, Du2024, zhang2025moiré-128}. By inducing a long-wavelength potential on top of the atomic lattice, one can induce band folding, reduce bandwidths, and strengthen electron-electron interactions~\cite{zhang2025moiré-128,li2023Wigner, Zhang2024Engineering, Nuckolls2024A}. A particularly successful implementation of this concept uses moiré patterns generated by twisting atomically thin layers~\cite{Cao2018correlated,guo2025superconductivity-263}. 
%
%In such systems, the moiré potential can dramatically suppress the kinetic energy of electrons and promote strong electron-electron interactions~\cite{zhang2025moiré-128,li2023Wigner, Zhang2024Engineering, Nuckolls2024A}. 
This strategy has led to the discovery of many novel phenomena, including correlated insulating states~\cite{Cao2018correlated,tian2024dominant-b2e},  unconventional superconductivity~\cite{Cao2018Unconventional,balents2020superconductivity,tanaka2025superfluid-65f}, topological states~\cite{chen2020tunable-f63,su2025moirédriven-896}, the fractional quantum Hall effect~\cite{lu2024fractional-60e,nuckolls2025spectroscopy-726}, and emergent magnetism~\cite{Sharpe2019Emergent,zhao2024emergence-34c}.  Notably, many of these phenomena involve spontaneous symmetry breaking, highlighting the connection between superlattice geometry and the emergence of new quantum phases.
\par
An alternative approach to engineer superlattices is to use the intrinsic electronic order in one material, such as charge density waves (CDWs), to impose a superlattice potential in an adjacent layer through proximity effects~\cite{tilak2024proximity,kim2024effect-e04,zhao2022moiré-82f}. 
Unlike moiré engineering via twist angles, where achieving the desired twist angle during fabrication can be challenging, 
%and the resulting superlattices are prone to crystal relaxations, 
superlattices based on intrinsic order offer an inherent advantage: 
%1) The superlattice period is independent of the twist angle, and 2)
when the two constituent lattices are nearly commensurate, the system spontaneously relaxes into an energetically favourable registry~\cite{woods2014commensurateincommensurate-d0c,yoo2019atomic-562,lai2025moire-16a}.
This `lock-in' mechanism, in which the layers self-align through slight rotations or lateral sliding, yields structures that are both energetically stable and experimentally reproducible.
As a result, this type of superlattice engineering provides a robust and deterministic platform for realising novel quantum phases. 
%overcoming the angle-precision limitations of  moiré systems. %\KJ{Lennart: I suggest to just remove the last (half) sentence "overcoming...", because (i) the twist angle precision in graphene as well as WSe2, MoTe2, etc seems to be pretty high and (ii) twisted structures also have such lock-in mechanisms, i.e., local minima in the free energy as function of twist angle.}

%the two constituent layers will slightly twist or slide into an energy-favourable stacking registry, a process that can cause the layers to "lock in" to the target angle, making these structures more energetically stable and experimentally accessible. This provides a powerful and potentially more reliable platform for engineering new quantum phases.

%an attractive solution arises when the two lattices are commensurate or nearly commensurate at specific angles ~\cite{gruner2018density}.  During the stacking of the vdW heterostructures, 

\par
%For superlattices imposed on graphene that respect its threefold rotational symmetry (\C), the geometry dictates how the electronic bands are folded into the new mini-Brillouin Zone (mBZ). Two primary folding scenarios exist. The first scenario maps the Dirac cones from graphene's original K and K' valleys onto the $\Gamma$ point of the mBZ~\cite{Bao2021Experimental}. This allows the superlattice potential to mediate intervalley scattering and open a band gap. In the second scenario, the valleys are folded to the corners of the mBZ (the K/K' points), where they remain distinct~\cite{hidalgo2023interlayer}. Here, the potential induces only intra-valley coupling. This breaks the emergent continuous rotational symmetry of the Dirac cones down to the discrete \C symmetry, producing effects such as trigonal warping rather than a full gap~\cite{ortix2011graphene-130}. The folding pathway is therefore a key determinant of the emergent electronic properties

%In graphene, the symmetry of the superlattice defines how the Dirac cones are reorganised in the mini-Brillouin zone (mBZ)~\cite{wallbank2013generic-397}.  
In graphene, the symmetry of the superlattice defines at which points the Dirac cones appear in the mini-Brillouin zone (mBZ)~\cite{wallbank2013generic-397}.  
As the K and $\mathrm{K}'$ valleys of graphene are related by time-reversal symmetry,
%and are located at high-symmetry points of the hexagonal Brillouin zone, 
there are only two  distinct folding configurations.  
In one configuration, both K and $\mathrm{K}'$ valleys are folded onto the $\Gamma$ point of the mBZ ($\Gamma$-folding), merging the two Dirac cones in momentum space~\cite{Bao2021Experimental,qiu2025giant-d55}. In the
other
%second (\KJ{replace second by other? or we replace the one configuration with the first configuration}) 
configuration, the valley identities are retained, with K and $\mathrm{K}'$ mapped separately to the  $\mathrm{K}_\mathrm{m}$ and $\mathrm{K}'_\mathrm{m}$ corners of the mBZ (K-folding)~\cite{ortix2011graphene-130,wallbank2013generic-397}. 
%The ability to achieve these two  different folding configurations represents a benchmark for engineering electronic states in the vdW heterostructures.
\par
While symmetry arguments predict these two folding scenarios, they do not completely determine the resulting electronic structure. 
Hybridisation with substrate bands, interlayer relaxation, and subtle variations in local stacking registry may profoundly reshape the band structure in ways that cannot be anticipated from geometry alone~\cite{yoo2019atomic-562,barré2024engineering-b09}.
Moreover, such effects may differ qualitatively between the two configurations: do $\Gamma$-folding and K-folding %($\mathrm{K}'$) 
produce equivalent electronic reconstructions, or does one geometry harbour instabilities absent in the other? 
Answering this question requires the combination of experiment and theory within a controlled material system.
%Answering this question requires the combination of experimental realisation and density functional theory (DFT) calculations of both configurations within a controlled material system.

Here, we demonstrate deterministic control over the Dirac cone folding by using the CDW in 1T-\NbSe~\cite{liu2021monolayer-ebe,nakata2021robust-bae,huang2025doped-cf4} to engineer two distinct, nearly commensurate superlattices on a graphene layer. 
%By changing the twist angle, 
We create a \Devi superlattice relative to the CDW lattice that exhibits $\Gamma$-folding, and a \Devii superlattice that exhibits K-folding. 
Using scanning tunnelling microscopy and spectroscopy (STM/STS), we show that the $\Gamma$-folded system preserves all underlying symmetries, while the K-folded system exhibits a spontaneous breaking of the \C symmetry.
%, which we corroborate by DFT calculations. 
%This symmetry breaking is not electronically driven, but originates from a subtle effect driven by the local hybridisation between the two layers, which is in turn caused by the varying stacking of graphene and \NbSe. 
This symmetry breaking is not electronically driven but originates from variations in the local stacking registry between graphene and \NbSe, which modulate the interlayer hybridisation. %A geometric interlayer interaction model confirms that  the interlayer coupling in the \Devii geometry is more sensitive to stacking variations and thus  prone to  breaking rotational symmetry.
%sliding and thus easily breaks the rotational symmetry. 
%This work presents a new strategy for building (near) commensurate superlattices by which control over band-folding pathways is achieved and uncovers a structurally driven mechanism for symmetry breaking.
%This work presents a new strategy for controlling band-folding pathways by using  superlattice engineering and uncovers a structurally driven mechanism for symmetry breaking.
% \clearpage
%===============================================================================
%\section*{Results}

\begin{figure}[p!]
     %\centering    
     
\includegraphics[width=\linewidth]{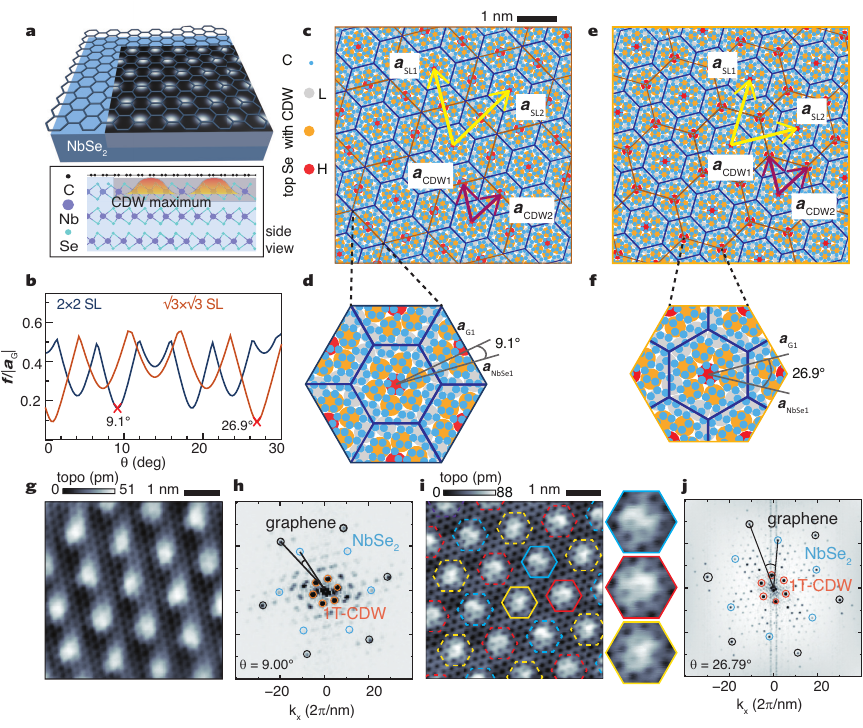
}
     \caption{
     \textbf{Superlattice Engineering.}
      \textbf{a,} Scheme of the graphene on 1T-\NbSe/2H-\NbSe van der Waals heterostructure. The side view illustrates the charge density wave (CDW) maxima. 
      \textbf{b,}  Twist-angle $\theta$ dependent mismatch between the graphene lattice and the \Devi (blue) and  \Devii (orange) superlattice of the 1T-CDW in units of the graphene lattice constant. Here, we use the Wood's notation \Devii to denote a superlattice 
with unit vectors $\sqrt{3}$ times larger than those of the CDW lattice, 
rotated by $30^{\circ}$~\cite{wood1964vocabulary-bfd}.
      \textbf{c,}~Scheme of the stacking configuration of a \Devi commensurate superlattice realised by compressing the graphene lattice by a heterostrain of $\varepsilon=4.1\%$.
      \textbf{d,}~Zoom-in of the supercell.
      \textbf{e, f,}~Same as (\textbf{c}, \textbf{d}) but for a \Devii commensurate superlattice with $\varepsilon=1.8\%$.
          %. The graphene lattice constant is suppressed by \SI{1.8}{\percent}.
        \textbf{g, h,} Atomically resolved STM topography and its FFT with $\theta=\SI{9.0}{\degree}$ and $\varepsilon=0.3\%$ ($V =\SI{0.5}{\volt}$, $I=\SI{200}{\pico\ampere}$).
        %The twist angle between graphene and \NbSe is determined to \SI{9.0}{\degree} following the method of ref.~\cite{Kerelsky2019Maximized}.
        \textbf{i, j,}  Same as (g, h) for the sample with $\theta=26.8^{\circ}$ and $\varepsilon=0.4\%$
        ($V =\SI{0.3}{\volt}$, $I=\SI{60}{\pico\ampere}$). Colored hexagons mark the 3 different CDW maxima which repeat periodically, building up the \Devii supercell.
          %\textbf{j,} FFT image of the topography shown in panel \textbf{i}.
   }
     \label{fig:fig1}
 \end{figure}

%===============================================================================
Through mechanical stacking with carefully selected rotation angles $\theta$, we create near-commensurate alignments between the \SoD CDW in 1T-\NbSe and a monolayer of graphene (Fig.~\ref{fig:fig1}a and Methods\ref{sec:method}).
% LK: The \ref{} on the methods section is broken!
%The $\sqrt{13}\times\!\sqrt{13}R\SI{13.9}{\degree}$ CDW of 1T-\NbSe provides a well-defined periodic potential. 
%Through mechanical stacking with carefully selected angles $\theta$, we create near-commensurate alignments between the CDW and a monolayer of graphene (Fig.~\ref{fig:fig1}a and Methods\ref{sec:method}). 
The condition for an exact commensurate superlattice is given by the linear Diophantine equation~\cite{perkins2025symmetry-032,zeller2014what-9df}:
\begin{equation} m\mathbf{R}(\theta)\boldsymbol{a}_{\rm G1} + n\mathbf{R}(\theta)\boldsymbol{a}_{\rm G2} = j\boldsymbol{a}_{\rm CDW1} + k\boldsymbol{a}_{\rm CDW2},\label{equa:equa1} 
\end{equation}
with $\boldsymbol{a}_{\rm G1}$ and  $\boldsymbol{a}_{\rm G2}$ as the primitive lattice vectors of graphene, $\boldsymbol{a}_{\rm CDW1}$ and  $\boldsymbol{a}_{\rm CDW2}$ as those of the CDW in 1T-\NbSe, and $\mathbf{R}(\theta)$ as the rotation of the graphene lattice with respect to the CDW by an angle $\theta$.
%Here, $\boldsymbol{a}_{\rm G1}$ and  $\boldsymbol{a}_{\rm G2}$ are the primitive lattice vectors of graphene, while $\boldsymbol{a}_{\rm CDW1}$ and  $\boldsymbol{a}_{\rm CDW2}$ correspond to  the lattice vectors of the CDW modulation in the 1T-\NbSe and $\mathbf{R}(\theta)$ represents the rotation of  the graphene lattice with respect to the CDW by an angle $\theta$. 
The integers $m,n,j$, and $k$ define the commensurate condition. %Since the graphene lattice constant ($|\boldsymbol{a}_{\rm G1}| = \SI{0.24}{\nano\meter}$) is much smaller than the CDW periodicity ($|\boldsymbol{a}_{\rm CDW1}| = \SI{1.24}{\nano\meter}$), we focus on the lower-order combinations of $j$ and $k$, which result in superlattices with experimentally accessible periodicities.

However, the different periodicities of the 1T-\NbSe CDW and the graphene lattice mean that equation~\ref{equa:equa1} has no integer solution. %However, due to the different periodicity of the 1T-\NbSe CDW and the graphene lattice, equation~\ref{equa:equa1} has no integer solution. 
%
%Due to the difference between the periodicity of  the 1T-\NbSe CDW and the graphene lattice constant, equation \ref{equa:equa1} has no integer solution. %For searching the near commensurate superstructure, we define the following function to quantify the degree of incommensurability:
This geometric incommensurability motivates a mismatch function that quantifies how closely the two lattices can approach commensurability:
%This geometric incommensurability motivates the definition of a mismatch function that quantifies how closely the two lattices can approach commensurability:
 %
%\begin{equation}
%f(\theta) = \min_{m,n\in \mathbb{Z}} %|m\boldsymbol{R}(\theta)\boldsymbol{a}_{\rm G1} + n\boldsymbol{R}(\theta)\boldsymbol{a}_{\rm G2} - j\boldsymbol{a}_{\rm CDW1}- k\boldsymbol{a}_{\rm CDW2}|. 
%\label{equa2:equa2}
%\end{equation}
%
\begin{equation}
f\left(\theta\right)
=
\min_{\substack{m,n \in \mathbb Z}}
\left|
m\,\mathbf{R}(\theta)\,\boldsymbol a_{\rm G1}
+
n\,\mathbf{R}(\theta)\,\boldsymbol a_{\rm G2}
-
j\,\boldsymbol a_{\rm CDW1}
-
k\,\boldsymbol a_{\rm CDW2}
\right| .
\label{equa2:equa2}
\end{equation}
This function computes the minimal distance between any rotated graphene lattice vector (spanned by the integers $m$ and $n$, and the rotation angle $\theta$) and the target CDW superlattice vector. 
The results %of equation \ref{equa2:equa2} 
for a \Devi superlattice (with $j = 2$ and $k = 0$) and a \Devii superlattice (with $j = k = 1$) are shown in Fig.~\ref{fig:fig1}b (for additional solutions see \Supple 1). 
These two geometries are of particular interest, because they produce the two distinct band-folding scenarios: the \Devi superlattice, whose stacking configuration is shown in Fig.~\ref{fig:fig1}c,d exhibits $\Gamma$-folding, while the \Devii superlattice (stacking configuration in Fig.~\ref{fig:fig1}e,f) shows K-folding, preserving valley identity.

Figures~\ref{fig:fig1}g-j show atomically resolved STM topographies and the corresponding fast Fourier transformations  (FFTs) of two samples with chosen $\theta$ close to those  for near-commensurate \Devi and \Devii superlattices.
Both topographic images reveal a strong periodic height modulation of the graphene layer, imprinted by the underlying triangular CDW lattice of the 1T-\NbSe. The CDW maxima, which correspond to the centres of the Star-of-David clusters characteristic of the 1T phase, appear bright in the topography.
%Both topographic images reveal a periodic apparent height  modulation of the graphene layer by the underlying triangular $\sqrt{13}\times\sqrt{13}R\SI{13.9}{\degree}$ CDW of the 1T-\NbSe. 
Following the procedure described in Ref.~\cite{Kerelsky2019Maximized}, we extract from the FFTs a very small heterostrain $\varepsilon<0.5\%$ in both samples (\Supple 2).
%the local stacking configurations at the CDW maxima appear nearly identical
Despite their similarities, the topographies of the two samples differ in a revealing way.  In the \Devi superlattice, all CDW maxima show  a nearly identical local contrast (Fig.~\ref{fig:fig1}g), whereas three distinct stacking configurations that repeat periodically across the surface can be identified in the \Devii system (Fig.~\ref{fig:fig1}i). 
%Based on the stacking geometry, these configurations correspond to (i) the A-sublattice carbon atom positioned above a CDW maximum, (ii) the B-sublattice carbon atom above a CDW maximum, and (iii) the hexagonal centre of graphene aligned with a CDW maximum (Fig.~\ref{fig:fig1}i).
This %observation 
suggests that in the \Devii superlattice, the precise location of the carbon atoms with respect to the CDW lattice has a stronger influence on the interactions between the two layers than in the \Devi superlattice.
As we will discuss below, this difference affects the electronic structure and the symmetries of the two samples.

\begin{figure}[ptb]
     \centering
     
\includegraphics[width=\linewidth]{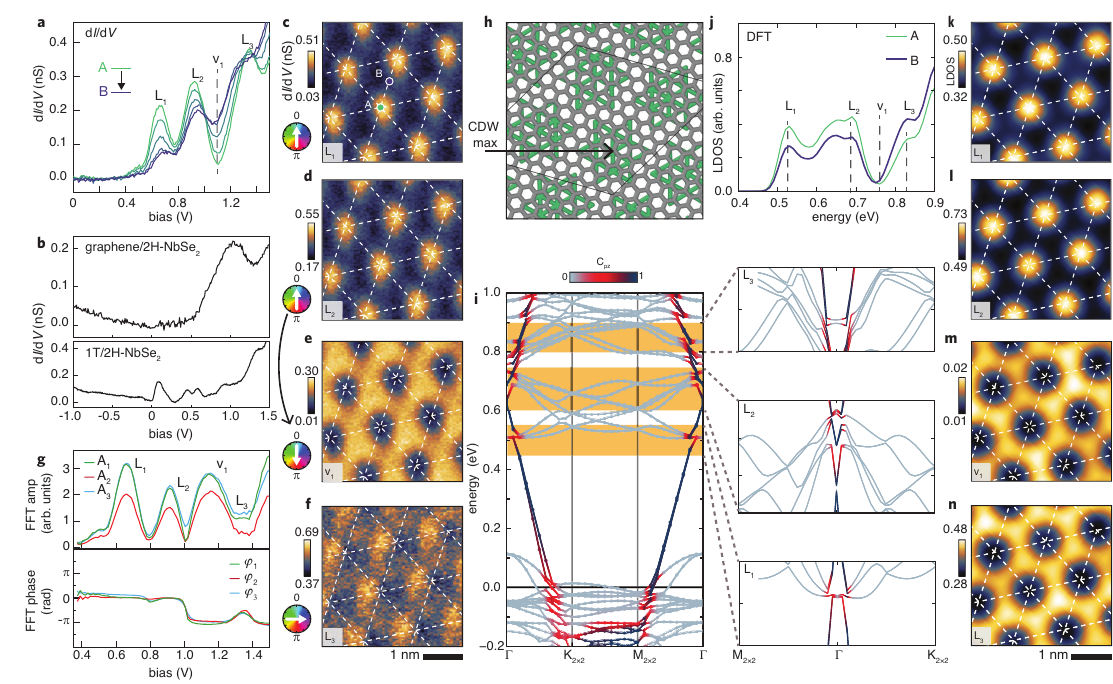}
     \caption{\textbf{Electronic structure of the $2\times 2$ superlattice.}
\textbf{a,}  
Spectra measured between the A- and B-sites, marked in  panel c,
show three peaks at energies $E_{\rm L1}= \SI{0.66(1)}{\electronvolt}$, $E_{\rm L2}= \SI{0.92(1)}{\electronvolt}$, and $E_{\rm L3}= \SI{1.36(1)}{\electronvolt}$ and a dip at $E_{\rm v1} = \SI{1.10(1)}{\electronvolt}$. 
\textbf{b,} On graphene/2H-\NbSe and 1T-\NbSe/2H-\NbSe these features are absent.
\textbf{c-f,}~\dIdV maps measured at biases corresponding to  
$E_{\rm L1-L3}$, and $E_{\rm v1}$.
%$L_{1-3}$, and $v_1$ in (a).
\textbf{g,} Evolution of the FFT amplitudes $A_{1-3}$ (top) and phases $\varphi_{1-3}$ (bottom) extracted from the \dIdV maps.
\textbf{h,}~Schematic  Model %of the \Devi graphene/1T-\NbSe superlattice 
used for the DFT calculations.
\textbf{i,}~Calculated band structure %along the high-symmetry directions 
of the superlattice mBZ and zoom-ins to the bands near the $L_{1-3}$ states.
The hybridised states are coded in red.
\textbf{j,}~Calculated LDOS at the A- and B-sites. %site) and  at the bridge-site
\textbf{k-n,} Calculated LDOS maps at energies corresponding to the $L_{1-3}$, and $v_1$ features.
} 
     \label{fig:fig2}
 \end{figure}

%In order to investigate the electronic properties of our samples, %the graphene/1T-\NbSe heterostructures, 
We begin by examining 
%We first  examine the electronic structure of 
the \Devi superlattice using differential conductance (\dIdV) spectroscopy, whose intensity is proportional to the local density of states (LDOS). %~\cite{tersoff1985theory-fa6}. 
%
%Figure~\ref{fig:fig1}g shows  an atomically resolved STM topography of the first Gr/1T-\NbSe sample. The image reveals the graphene layer periodically modulated by a triangular superlattice. The Fast Fourier Transform (FFT) analysis of the topography confirms that the  periodicity of the modulation within the graphene layer matches the $\sqrt{13}\times\sqrt{13}R\SI{13.9}{\degree}$ CDW phase  of the underlying 1T-\NbSe. Applying the   analysis method described in ref.~\cite{Kerelsky2019Maximized} to our system, we extract a twist angle \SI{9.0}{\degree}  between the graphene and \NbSe. This angle is in  close agreement  with the predicted value for the nearly commensurate $2\times 2$  superlattice configuration discussed earlier. 
%
%Likewise, the high-resolution STM topography of the second sample is shown in Fig.~\ref{fig:fig1}i.  Three circles highlight distinct local  stacking configurations of the graphene lattice atop the CDW maxima.  The observed superlattice modulation is consistent with a $\sqrt{3}\times\sqrt{3}R\SI{30}{\degree}$ configuration relative to the 1T-\NbSe. FFT analysis confirms the twist angle required to achieve the  near-commensurate $\sqrt{3}\times\sqrt{3}R\SI{30}{\degree}$ superlattice. 
%
%In order to investigate the electronic properties of our samples, %the graphene/1T-\NbSe heterostructures, we begin by examining the \Devi superlattice using differential conductance (\dIdV) spectroscopy, whose intensity is proportional to the local density of states (LDOS). 
Figure~\ref{fig:fig2}a shows \dIdV spectra acquired along a line, starting at the centre of the CDW maximum (referred to as A-site), and moving towards the midpoint between two neighbouring clusters (B-site). %, as  indicated in  Fig.~\ref{fig:fig2}c %(spectra of a wider bias range can be found in \Supple~3).
%\KJ{Over a wide bias range from -1.5 to \SI{1.5}{\volt}, the negative-bias side remains featureless, while all electronically relevant features appear at positive bias (\Supple 3); we therefore focus on this regime.}
At the A-site, the \dIdV spectrum exhibits three pronounced peaks at energies $E_{\rm L1}$, $E_{\rm L2}$, and $E_{\rm L3}$ (see \Supple 3 and 4).
%\KJ{with the positions determined by Gaussian fitting (\Supple 4).} 
These peaks are absent in measurements taken at 1T/2H-\NbSe or graphene/2H-\NbSe heterostructures (Fig.~\ref{fig:fig2}b), confirming that they arise specifically from the interaction between the graphene sheet and the 1T-\NbSe layer. 
%\KJ{That is from the superlattice-induced band reconstruction.}
%

As the tip moves from the A- to the B-site, the peaks decay in intensity, while a new 
feature at $E_{\rm v1}$ emerges with maximum intensity at the B-site.
%In contrast,  a distinct spectral feature at $E_{\rm v1}=(1.10\pm0.01)$~eV follows the opposite behaviour, with maximum intensity at the B-site. 
These complementary behaviours are visualised directly in constant energy \dIdV maps. % extracted from grid spectroscopy data %(see Supp.~Inf.~and Supp.~Fig.~X).
At $E_{\rm L1}$ and $E_{\rm L2}$, the \dIdV maps exhibit a dot-like structure with high intensity at the A-sites (Fig.~\ref{fig:fig2}c,d).
The FFTs of these maps reveal  significant intensity exclusively at the Bragg peaks corresponding to the reciprocal vectors of the CDW, $\boldsymbol{Q}_i^\text{CDW}$, suggesting that the spatial variation of the \dIdV can be described by a Fourier series expansion over these wavevectors as:
%
%\begin{equation}
    %\mathrm{d}I/\mathrm{d}V\left(\boldsymbol{r},V\right) = A_0\left(V\right)+ \sum_{i=1}^3 A_i\left(V\right) \cos\left(\boldsymbol{Q}_i^\text{CDW}\cdot \boldsymbol{r}+\varphi_i(V)\right).
    %\label{equ:FFT-analysis}
%\end{equation}
\begin{equation}
\mathrm{d}I/\mathrm{d}V\left(\boldsymbol{r},V\right)
=
A_0\left(V\right)
+
\sum_{i=1}^{3}
A_i\left(V\right)\,
\cos\!\left(
\boldsymbol{Q}_i^{\mathrm{CDW}} \!\cdot\! \boldsymbol{r}
+
\varphi_i\left(V\right)
\right).
\label{equ:FFT-analysis}
\end{equation}
In this equation, $A_{1,2,3}$ are the amplitudes and $\varphi_{1,2,3}$ the phase-shifts of the LDOS modulation along the three fundamental directions of the CDW~\cite{pasztor2019holographic-369} (see \Supple 5). 
%All $A_i$ %three amplitudes show similar behaviour, with maxima occurring at $E_{L1}$, $E_{L2}$, and $E_{v1}$ (Fig.~\ref{fig:fig2}g). 
%The minor difference in relative intensity might originate from a slight directional dependence of the probing tip.
%
The phases $\varphi_i (V)$ define the real-space registry of the LDOS modulation relative to the CDW superlattice. 
When $\varphi_i =0$, the \dIdV maxima align with the A-sites, producing the dot-like pattern as seen at $L_1$ and $L_2$. Near $V\approx \SI{1}{\volt}$, all three phases simultaneously undergo  a $\pi$-shift,  resulting in an inversion of the LDOS contrast and a transformation of the dot-like pattern into a honeycomb-like structure (Fig.~\ref{fig:fig2}e). At the energy of $L_3$, the phases adopt intermediate values, $\varphi_{1-3} \approx -\pi/2$, shifting the intensity maxima to positions between the A- and B-sites (Fig.~\ref{fig:fig2}f).
A significant finding of this analysis is that $A_{1-3}$ and $\varphi_{1-3}$ evolve similarly across the measured energy range (Fig.~\ref{fig:fig2}g), indicating the preservation of the underlying $C_3$ rotational symmetry of the CDW lattice. The small difference in relative intensity might originate from a slight directional dependence of the probing tip.

To understand the origin of the observed spectral features, we performed DFT calculations (see Methods) for a commensurate \Devi %graphene/1T-\NbSe 
superlattice (Fig.~\ref{fig:fig2}h). %The calculations successfully corroborate our experimental findings and reveal the underlying physical mechanism. 
The superlattice potential defines the mBZ into which the original bands are folded (Fig.~\ref{fig:fig2}i). %, with graphene-derived bands in dark blue, 1T-\NbSe bands in grey, and hybridised states in red. 
The key feature of this geometry is the $\Gamma$-folding: the Dirac cones from both the K and $\mathrm{K}'$ valleys of pristine graphene fold onto the $\Gamma$ point of the mBZ. %This valley degeneracy, combined with hybridisation with the \NbSe bands, produces  anticrossing  that  reconstructs the electronic structure. 
This valley degeneracy, combined with hybridisation with the \NbSe bands, produces peaks in the calculated LDOS (Fig.~\ref{fig:fig2}j) and the corresponding LDOS maps (Fig.~\ref{fig:fig2}k-n), which  qualitatively reproduce the experimental observations. %on a qualitative level, as expected for DFT on large supercells 

  \begin{figure}[pth]
     \centering
\includegraphics[width=\linewidth]{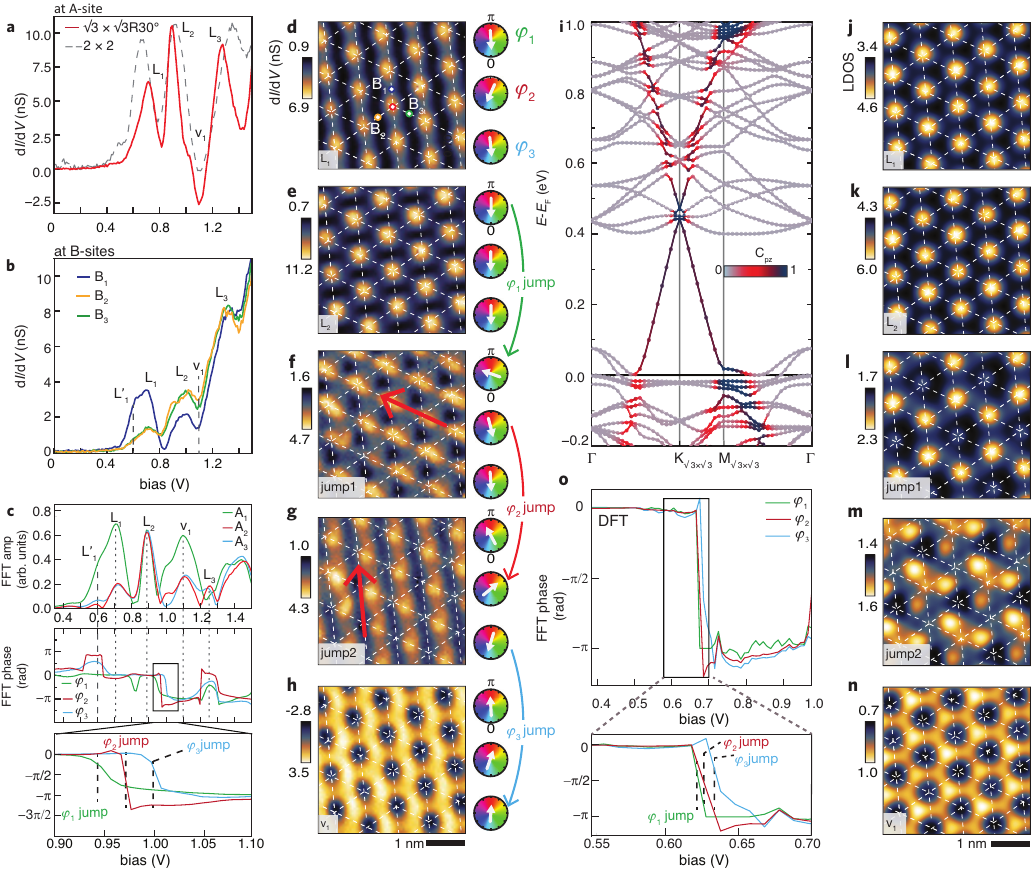}
     \caption{\textbf{Electronic structure of the $\sqrt{3}\times\!\sqrt{3}\ \mathrm{R}30^\circ$ superlattice}
\textbf{a,} Spectrum taken at the A-site (solid line)
%, \It = \SI{500}{\pico\ampere}, \Vb = \SI{1.5}{\volt}, and \Vmod = \SI{10}{\milli\volt}) 
show peaks at energies $E_{\rm L1}=0.71\pm0.01$~eV, $E_{\rm L2}=0.90\pm0.01$~eV, and $E_{\rm L3}=1.27\pm0.01$~eV and a dip at  $E_{\rm v1}=1.10\pm0.01$~eV. The spectrum measured in the $2\times 2$ superlattice (dashed line, $\times25$) is shown for comparison.
%to one measured in the $2\times 2$ superlattice (dashed line, $\times25$). \KJ{The spectrum shows three peaks $L_1$ at\SI{0.71(1)}{\volt}, $L_2$ at\SI{0.90(1)}{\volt} and $L_3$ at \SI{1.27(1)}{\volt} as well as a dip $v_1$ at \SI{1.10(1)}{\volt}, at comparable positions to those in the \Devi superlattice.}
\textbf{b,}  Spectra recorded at the CDW minima B$_1$ (blue), B$_2$ (orange) and B$_3$ (green), which are rotated around the CDW maximum by \SI{120}{\degree} with respect to each other. At B$_1$ an additional peak $L_1'$ appears at $E_{\rm L1'}=0.62\pm0.01$~eV.
\textbf{c,} Phase and amplitude from the FFT analysis of the \dIdV data as a function of bias.
\textbf{d-h,}  \dIdV maps at the indicated energies.
\textbf{i,} Calculated band structure of the \Devii 
%superlattice along the high-symmetry directions of 
superlattice mini-Brillouin zone.
\textbf{j-n,} DFT calculated LDOS maps, at the corresponding states.
\textbf{o,} Similar phase analysis using the DFT-calculated LDOS data. 
} 
     \label{fig:fig3}
 \end{figure}
%\textbf{k,} Atomic model of $\sqrt{3}\times \sqrt{3}$  commensurate superlattice used for the DFT calculations.

%
We now turn to the electronic structure of the \Devii superlattice. At the A-site, we observe spectral features  that are similar to those in the \Devi superlattice, albeit with a subtle shift of the peak positions (Fig.~\ref{fig:fig3}a). 
The  three B-sites, which are related by a \SI{120}{\degree} rotation around the A-site, however, are electronically inequivalent: The $\mathrm{B}_1$-site shows an additional peak ($L'_1$) and increased intensity at $L_1$, but decreased intensity at $L_2$, compared to the $\mathrm{B}_2$- and $\mathrm{B}_3$-sites (Fig.~\ref{fig:fig3}b).
 Applying the same FFT phase analysis to
 this
 %the \Devii 
 superlattice 
 reveals a much more complex behaviour (Fig.~\ref{fig:fig3}c):
At the energy of peak $L_1$, we observe a dot-like pattern, accompanied by stripes along one direction (Fig.~\ref{fig:fig3}d), resulting in amplitudes $A_1\approx A_3\ll A_2$ and nearly aligned phases $\left(\varphi_{1-3}\approx 0 \right)$, providing a clear evidence of broken \C symmetry. As the energy increases to $L_2$, the 
%this weakly broken 
\C symmetry is restored, yielding a lattice of symmetric circular dots at A-sites  (Fig.~\ref{fig:fig3}e). %\KJ{Extended STS grid maps, FFT amplitudes, and FFT phases for the \Devii sample are presented in \Supple 6.}
%
%
%Between $L_2$ and $L_3$, at the spectroscopic feature $v_1$, the LDOS pattern evolves into a honeycomb-like structure by  inverting the intensitI/y contrast. % ($\varphi_1\approx \varphi_2 \approx\varphi_3\approx -\pi$), corresponding to a $\pi$-phase  jump . 
%This transition from circular dots ($\varphi_i \approx 0$)  to honeycomb ($\varphi_i \approx -\pi$) occurs through a series of staggered phase jumps ( see the inserted panel of Fig.~\ref{fig:fig3}c).  Unlike in \Devi, where the three phases jump coincidentally at the same bias, the transitions of three directions in \Devii occur at slightly different biases, thus this transition breaks \C rotational symmetry and leads to an asymmetric intermediate pattern. The phase $\varphi_2$ starts its transition at the lowest bias, distorting the circular dot pattern into an elliptical shape (Fig.~\ref{fig:fig3}g).  Subsequently, $\varphi_3$ jumps to $-\pi$, resulting in a stripe pattern. This strip pattern is rotated by \SI{60}{\degree} relative to the previous pattern (Fig.~\ref{fig:fig3}h). Finally, $\varphi_1$ undergoes the phase jump at the highest bias, completing the transformation from the dot-pattern into the honeycomb pattern.  These staggered jumps correspond to the three distinct local minima observed in the amplitudes $A_q(V)$ within the $V_1$ bias range. After the jumps, the amplitudes remain unequal, with $A_2$ remaining dominant   $\left(A_2> A_1\approx A_3\right)$.

Between $L_2$ and $v_1$, we again observe a transition from a dot-like to a honeycomb-like LDOS pattern, but, unlike the symmetry-preserving transition in the \Devi system, this transition proceeds through a series of staggered phase jumps that progressively break \C symmetry (Fig.~\ref{fig:fig3}c). 
The first phase jump distorts the dots into ellipses along one axis (Fig.~\ref{fig:fig3}f); the second phase jump creates a transient stripe pattern (Fig.~\ref{fig:fig3}g); and the third phase jump completes the transformation to the honeycomb pattern (Fig.~\ref{fig:fig3}h). However, even at $v_1$, the amplitudes remain unequal, with $A_2$ being dominant, confirming the persistence of the broken symmetry. 
%\KJ{This  finding is unexpected as the \Devii superlattice nominally preserves \C symmetry.}
%I take it out here. Let's discuss it below.

%As a higher bias, another set of phase jumps  occurs near $L_3$ at \SI{1.2}{\volt}, transmitting the phase to $\varphi_1 \approx -\pi/4,\varphi_2\approx -\pi/2,\varphi_3\approx0$, corresponding to a complicated interference pattern (Fig.~\ref{fig:fig3}j). As further bias increases above \SI{1.3}{\volt}, the phases re-align to $\varphi_1 \approx -\pi, \varphi_2 \approx -\pi, \varphi_3 \approx -\pi$, and the amplitudes converge again, restoring a  honeycomb pattern preserving \C rotational symmetry. This behaviour is a hallmark of a symmetry-broken  state. 

DFT calculations on this %\Devii commensurate 
supercell confirm the K-folding (Fig.~\ref{fig:fig3}i). Hybridisation with the 1T-\NbSe bands gives rise to a series of prominent peaks, and the calculated LDOS maps qualitatively reproduce  the measured \dIdV maps (Fig.~\ref{fig:fig3}j-n). Notably, the same FFT phase analysis applied to the calculated maps reveals a comparable series of staggered phase jumps (Fig.~\ref{fig:fig3}o), yielding asymmetric intermediate states with broken \C symmetry (Fig.~\ref{fig:fig3}m). 

  \begin{figure}[t]
     \centering
    \includegraphics{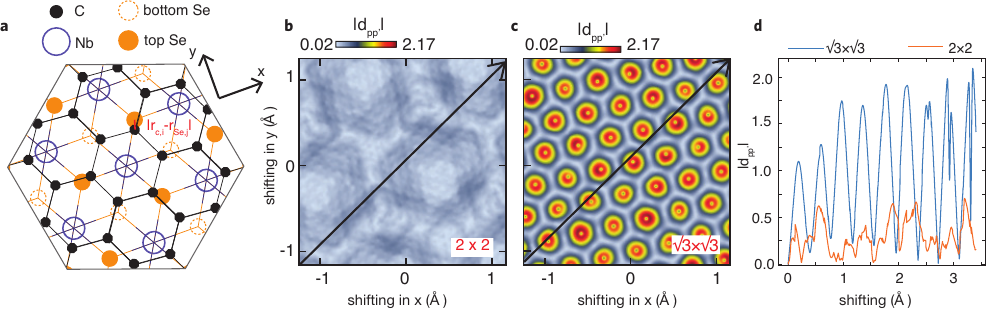}
     \caption{\textbf{Geometric origin of the symmetry breaking.}
\textbf{a,} Schematic of the interlayer interaction model.
\textbf{b,} 2D resolved sliding map of the calculated hybridisation magnitude $|d_{pp}|$ for the \Devi superlattice with sliding directions defined in \textbf{a}.
\textbf{c,} 2D resolved sliding map of the $|d_{pp}|$ for the \Devii superlattice.
%Comparison of the Direction-resolved spatial average of the hybridisation between the \Devi and \Devii superlattices for the commensurate configurations.
\textbf{d,}  Comparison of the  $|d_{pp}|$ as a function of sliding between these two commensurate configurations, along the sliding paths indicated in \textbf{b} and \textbf{c}. \label{fig:fig4}}
 \end{figure}
 
Although the DFT calculations capture the broken symmetry in the \Devii superlattice and its absence in the \Devi superlattice, the microscopic origin of this symmetry breaking remains unclear within DFT. To elucidate the mechanism, we developed a simplified atomistic model that quantifies the spatial average of the interlayer coupling based on local atomic registry. This model enables us to simulate the influence of lateral sliding of the graphene layer with respect to the \NbSe and the influence of isotropic strain on the interlayer hybridisation.

%Having established that DFT reproduces the experimental observation in symmetry breaking in \Devii superlattice, we now address its microscopic origin by comparing the hybridisation in \Devi and \Devii superlattices. We developed a geometric model that quantifies the directional-resolved spatial average of interlayer coupling based on local atomic registry. %For each carbon atom, we compute a unit vector pointing toward its nearest selenium neighbour in the 1T-\NbSe layer, weighted by an exponentially decaying function of their separation (motivated by the distance dependence of Slater-Koster hopping integrals). 

%First, we evaluated the commensurate configurations of \Devi and \Devii superlattices. As shown in Fig.\ref{fig:4}a,
%a spontaneous in-plane shift of the graphene layer relative to the underlying CDW. To gain physical intuition for why the \Devii superlattice is susceptible to such a relaxation while the \Devi is not, we developed a toy model to quantify the real-space interlayer interactions based on local atomic registry.

Our model quantifies the net directional preference of hybridisation between the $p_z$ orbitals of the carbon atoms in the graphene layer and the $p$ orbitals of the  topmost Se atoms in the \NbSe layer (Fig.~\ref{fig:fig4}a):
\begin{equation}
    \bvec d_{pp} = 
    \sum_{i,j} 
    \frac{\bvec r_{\mathrm{C},i} - \bvec r_{\mathrm{Se},j}}
    {\left|\bvec r_{\mathrm{C},i} - \bvec r_{\mathrm{Se},j}\right|} 
    w\!\left(\left|\bvec r_{\mathrm{C},i} - \bvec r_{\mathrm{Se},j}\right|\right)\,,
\end{equation}
where the sum runs over all C atoms of the graphene %$i$
and all Se atoms %$j$ 
in the topmost \NbSe layer and $\bvec r_{\mathrm{C},i}$ ($\bvec r_{\mathrm{Se},j}$) denotes the position of the $i$-th carbon ($j$-th selenium) atom.
%The model averages over the weighted unit vectors pointing from each carbon atom to its nearest selenium neighbour. \KJ{Here, $\bvec r_{\mathrm{C},i}$ ($\bvec r_{\mathrm{Se},j}$) denotes the position of the i-th carbon (j-th selenium) atom.} 
%Here, $\bvec r_{\mathrm{C},i}$ is the position of the $i$-th carbon atom and $\bvec r_{\mathrm{Se},i}$ is that of its nearest Se neighbour. 
The exponential weighting function $w(r) = e^{-\alpha r}$ represents  the distance dependence of Slater-Koster tight-binding parameters arising from orbital overlap~\cite{slater1954simplified-305}. The decay constant  $\alpha=\SI{0.3}{\per\angstrom}$ ensures that nearest-neighbour C–Se pairs dominate the hybridisation. Further details  are provided in \Supple 6. 

%The model quantifies the net directional preference for the graphene layer by calculating a  hybridisation vector, $\bvec d_{pp}$. This vector sums the unit vectors pointing from each carbon atom to its nearest selenium neighbour in the 1T-\NbSe layer, weighted by their separation:

%where $\bvec r_{\mathrm{C},i}$ is the position of the $i$-th carbon atom and $\bvec r_{\mathrm{Se},i}$ is that of its nearest Se neighbour. Motivated by the distance dependence of Slater-Koster tight-binding parameters, we use an exponential weighting function $w(r) = e^{-\alpha r}$ (with heuristic parameter $\alpha=0.3\,\text{\AA}^{-1}$) to ensure that closer atom pairs contribute more strongly. The detailed analysis can be found in the SI. % The out-of-plane buckling of the Se atoms due to the CDW is included by modulating their $z$-positions as
%\begin{equation}
%    z(\bvec r_\mathrm{2D}) = \frac29 \bigg[\frac32 + \sum_{i\in\{1,2,3\}} \cos(\bvec r_\mathrm{2D} \cdot \bvec G_i) \bigg]\,,
%\end{equation}

The results for laterally sliding graphene in the commensurate \Devi and \Devii superlattices are shown in Fig.~\ref{fig:fig4}b-d, and reveal a striking difference. In the \Devi superlattice, $\bvec d_{pp}$ remains nearly constant under sliding, indicating a flat hybridisation landscape that is insensitive to the local registry. In the \Devii  superlattice, by contrast, the average hybridisation is strongly modulated. This behaviour persists when we allow for an incommensurate graphene lattice (\Supple 7). %\KJ{ The strong sliding-dependent hybridisation of the \Devii geometry implies that a slight lateral shift can lift the degeneracy among the three CDW directions, providing a mechanism for the observed \C symmetry breaking.}  %These reveal a surprising difference between the two systems. While $|\boldsymbol{d}_{pp}|$ in the \Devi superlattice remains relatively constant, we observe a strong modulation of the average hybridisation in the \Devii superlattice.Similar behaviour occurs when we allow for an incommensurate graphene lattice (\Supple~7).
%===============================================================================
%\section*{Discussion}
%===============================================================================

%reveal that the hybridisation strength  is  different in the two geometries. For the ideal commensurate superlattices of \Devi and \Devii,  the \Devii superlattice displays a strong modulated hybridisation landscape with pronounced peaks and valleys, whereas the \Devi superlattice shows a comparatively smooth spatial profile (Fig.~\ref{fig:fig4}a).  When we introduce   lattice scaling to model the near-commensurate conditions of our experimental samples,  the two configurations respond in different ways (Fig.~\ref{fig:fig4}b). The hybridisation modulation in  \Devii geometry is strongly suppressed, indicating that small deviations from ideal commensuration dramatically redistribute the interlayer coupling.  In contrast, the  modulation in \Devi remains largely unchanged (Fig.~\ref{fig:fig4}b).

This different response enables us to address the origin of the symmetry breaking. In the \Devii geometry, the hybridisation landscape of the ideal structure is highly susceptible to perturbations; even a small lateral shift or strain can break the delicate balance of the three equivalent CDW directions and lift their degeneracy. The \Devi geometry, with its smoother hybridisation landscape, is more robust against such perturbations and maintains \C symmetry even under realistic sample conditions. Therefore, the symmetry breaking observed in the \Devii superlattice  arises from a geometric instability that is intrinsic to this superlattice configuration.

%This system is expected to be insensitive to lateral shifting, which is in agreement with our experimental results.
%(Fig.~\ref{fig:fig4}c and d). However, the change between the high symmetry stacking configuration and the shifting configuration is not very large.  
\par
%The situation is dramatically different for the \Devii superlattice. Here, the hybridisation in the high-symmetry configuration is strongly modulated, forming a sharp, dot-like pattern where dots are separated by only \SI{0.38}{\angstrom} (Fig.~\ref{fig:fig4}e, f). This highly structured hybridisation is remarkably sensitive to lateral shifts. As shown in Fig.~\ref{fig:fig4}g and h, even a small displacement of the graphene layer causes this dot-like pattern to smear out, indicating a rapid change in the interlayer coupling.

The superlattice engineering strategy demonstrated here can be naturally extended to a broad class of vdW heterostructures beyond our specific example.
%of graphene/1T‑NbSe$_2$. 
By selecting host materials with different CDW periodicities and symmetries, one can systematically design band‑folding pathways and target specific points of the mBZ for Dirac cones or other band features. 
Promising candidates include graphene on 1T‑TaS$_2$~\cite{tilak2024proximity}, whose \SoD CDW has a slightly larger periodicity than that of 1T‑NbSe$_2$, as well as other transition‑metal dichalcogenides with commensurate or nearly commensurate CDWs. 
In such systems, the competition between different %nearly commensurate 
registries, coupled to the structural instabilities revealed in this work, is expected to generate a rich landscape of symmetry‑broken ground states, including nematic and ferroic‑like orders that could be tuned by strain, pressure, or electrostatic gating. %, or substrate choice.

Our approach can be extended to semiconducting 2D crystals such as WSe$_2$ on 1T‑\NbSe or 1T‑TaS$_2$. In these heterostructures, CDW‑induced superlattices provide a route to engineer not only electronic minibands but also excitonic dispersions, valley selection rules, and radiative lifetimes through zone folding of exciton states~\cite{oudich2024engineered-e09}. This opens opportunities to realise designer excitonic lattices, control interlayer and moiré‑like excitons \cite{torre2025advanced-b8f}. More broadly, combining CDW‑based proximity superlattices with twist, strain, or multilayer stacking could enable hybrid CDW-moiré platforms in which structural relaxation, electronic correlations, and topology are all on comparable energy scales. 
Our results thus establish CDW‑mediated superlattice engineering as a robust and deterministic strategy for creating and controlling emergent quantum phases in vdW heterostructures, and motivate future experiments that exploit structural degrees of freedom as an equal partner to electronic interactions in the design of quantum materials.

\clearpage

%\begin{figure}[h]
%    \centering
%    \includegraphics[width=1\linewidth]{Figures/Fig5/fig5_v2.eps}
%    \caption{\textbf{Potential superlattice engineering.}
%     \textbf{a,}  Twist-angle $\theta$ dependent mismatch between the graphene lattice and the $1\times 1$ (yellow), $\sqrt{3}\times \sqrt{3}R30^\circ$ (green) and $2\times2$ superlattices of the 1T-TaS$_2$ CDW in units of the graphene lattice constant.
%    \textbf{b,} Twist-angle $\theta$ dependent mismatch between the WSe$_2$ lattice and the $1\times 1$ (yellow), $\sqrt{3}\times \sqrt{3}R30^\circ$ (green) and $2\times2$ superlattices of the 1T-\NbSe CDW in units of the  WSe$_2$ lattice constant.
%  \textbf{c,} Twist-angle $\theta$ dependent mismatch between the WSe$_2$ lattice and the $1\times 1$ (yellow), $\sqrt{3}\times \sqrt{3}R30^\circ$ (green) and $2\times2$ superlattices of the 1T-TaS$_2$ CDW in units of the  WSe$_2$ lattice constant.
%    }
%    \label{fig:fig5}
%\end{figure}
%\clearpage

%===============================================================================
\section*{Methods \label{sec:method}}
%===============================================================================

\subsection{Sample fabrication}
We fabricated the  heterostructures in an Ar-filled glove box following a suspended dry pickup and flip-over assembly technique~\cite{Jin2023}. We then placed the assembled heterostructure on top of a \SI{285}{\nano\meter} Si/SiO$_2$ chip on which we constructed contact and tip guidance electrodes~\cite{Martinez-Castro2018} by electron beam lithography (\Supple 8). %Finally, we transferred the samples to the STM using an ultra-high vacuum suitcase with a base pressure of $\approx\SI{1e-8}{\milli\bar}$.

Initial STM topographies of the graphene/2H-\NbSe heterostructures resolve the graphene lattice, the underlying \NbSe lattice, and the resulting moiré superlattice, but show no signatures of the $3\times3$ charge-density-wave (CDW) order (\Supple 9), consistent with previous reports~\cite{chen2020visualizing-890}. To realise graphene on 1T-\NbSe, we employ local bias pulsing with the STM tip on graphene/2H-\NbSe~\cite{bischoff2017nanoscale-01b}, which induces a controlled phase transition from the 2H to the 1T polytype in the \NbSe layer beneath the graphene (\Supple 10).

\subsection{STM and STS measurements}
%\MT{All measurements were performed in a commercial Sigma Polar instrument at $T= \SI{6}{\kelvin}$ in ultra-high vacuum ($\leq\SI{1e-10}{\milli\bar}$)}. 
All measurements were performed in ultra-high vacuum ($\leq\SI{1e-10}{\milli\bar}$) and at a base temperature of $T= \SI{6}{\kelvin}$.
As tip we used electrochemically etched tungsten tips, flashed in ultra-high vacuum to remove the oxide layer and contamination. 
%To ensure a clean tunnelling junction, we checked the W tips on an Ag(111) crystal. 
Additionally, the tip was prepared in-situ by repeatedly indenting it into the deposited gold contacts until it showed a flat density of states. STS was performed using a standard lock-in technique at a modulation frequency of \SI{877}{\hertz} and modulation voltages $V_{\rm mod}=10$~mV.

%===============================================================================
\subsection{DFT calculations}
%===============================================================================
\label{sec:methods-dft}

Density functional theory calculations were performed in a plane-wave pseudopotential framework. We first relaxed the \SoD\ CDW unit cell of 1T-\NbSe with fixed in-plane lattice parameters, and then constructed commensurate \Devi\ and \Devii\ graphene/1T-\NbSe supercells by  straining graphene homoaxially and twisting its lattice vectors. The graphene layer was initially placed $\sim 5$~\AA\ above the Nb plane, and all atomic positions in the heterostructures were relaxed.

We used a generalized-gradient exchange–correlation functional with van der Waals corrections, ultrasoft pseudopotentials, a vacuum spacing of 20\,Å, and kinetic-energy cutoffs of 40/400\,Ry for wavefunctions/charge density. Moderately dense \(k\)-point meshes were employed for self-consistent calculations and further refined for non-self-consistent band-structure, (projected) DOS and real-space LDOS calculations. The resulting electronic structures capture the key qualitative differences between the \Devi\ and \Devii\ superlattices, including their distinct band folding and symmetry properties.
More details can be found in the \Supple 11.

\section*{Data availability}
%===============================================================================

All data presented in this study are available on the Jülich Data Repository at xxxxxx

\bibliography{ref}

@article{Jin2023, 
year = {2023}, 
volume = {11},
number = {1},
pages = {2300658},
title = {{Assembly of Arbitrary Designer Heterostructures with Atomically Clean Interfaces}}, 
author = {Jin, Keda and Wichmann, Tobias and Wenzel, Sabine and Samuely, Tomas and Onufriienko, Oleksander and Szabó, Pavol and Watanabe, Kenji and Taniguchi, Takashi and Yan, Jiaqiang and Tautz, F. Stefan and Lüpke, Felix and Ternes, Markus and Martinez‐Castro, Jose}, 
journal = {Adv. Mater. Interfaces}, 
doi = {10.1002/admi.202300658}, 
}

@article{perkins2025symmetry-032, 
  year     = {2025}, 
  title    = {Symmetry preservation in commensurate twisted bilayers}, 
  author   = {Perkins, David T. S.}, 
  journal  = {Phys. Rev. B}, 
  issn     = {2469-9950}, 
  doi      = {10.1103/r61z-jhrn}, 
  pages    = {035410}, 
  number   = {3}, 
  volume   = {112}
}

@article{zeller2014what-9df, 
  year     = {2014}, 
  title    = {What are the possible moiré patterns of graphene on hexagonally packed surfaces? {Universal} solution for hexagonal coincidence lattices, derived by a geometric construction}, 
  author   = {Zeller, Patrick and Günther, Sebastian}, 
  journal  = {New J. Phys.}, 
  doi      = {10.1088/1367-2630/16/8/083028}, 
  pages    = {083028}, 
  number   = {8}, 
  volume   = {16}
}

@article{Martinez-Castro2018, 
year = {2018}, 
title = {{Scanning Tunneling Microscopy of an Air Sensitive Dichalcogenide Through an Encapsulating Layer}}, 
author = {Martinez-Castro, Jose and Mauro, Diego and Pásztor, \'{A}rp\'{a}d and Guti\'{e}rrez-Lezama, Ignacio and Scarfato, Alessandro and Morpurgo, Alberto F and Renner, Christoph}, 
journal = {Nano Lett.}, 
doi = {10.1021/acs.nanolett.8b01978}, 
pages = {6696--6702}, 
number = {11}, 
volume = {18}, 
}

@article{Kerelsky2019Maximized, 
year = {2019}, 
title = {{Maximized electron interactions at the magic angle in twisted bilayer graphene}}, 
author = {Kerelsky, Alexander and McGilly, Leo J. and Kennes, Dante M. and Xian, Lede and Yankowitz, Matthew and Chen, Shaowen and Watanabe, K. and Taniguchi, T. and Hone, James and Dean, Cory and Rubio, Angel and Pasupathy, Abhay N.}, 
journal = {Nature}, 
doi = {10.1038/s41586-019-1431-9}, 
pages = {95--100}, 
number = {7767}, 
volume = {572}, 
}

@article{Bistritzer2011moire, 
year = {2011}, 
title = {{Moiré bands in twisted double-layer graphene}}, 
author = {Bistritzer, Rafi and MacDonald, Allan H.}, 
journal = {Proc. Natl. Acad. Sci. U.S.A}, 
issn = {0027-8424}, 
doi = {10.1073/pnas.1108174108}, 
pages = {12233--12237}, 
number = {30}, 
volume = {108}
}

@article{Cao2018correlated, 
year = {2018}, 
title = {{Correlated insulator behaviour at half-filling in magic-angle graphene superlattices}}, 
author = {Cao, Yuan and Fatemi, Valla and Demir, Ahmet and Fang, Shiang and Tomarken, Spencer L. and Luo, Jason Y. and Sanchez-Yamagishi, Javier D. and Watanabe, Kenji and Taniguchi, Takashi and Kaxiras, Efthimios and Ashoori, Ray C. and Jarillo-Herrero, Pablo}, 
journal = {Nature}, 
issn = {0028-0836}, 
doi = {10.1038/nature26154}, 
pages = {80--84}, 
number = {7699}, 
volume = {556}
}

@article{Li2023Wigner, 
year = {2024}, 
title = {{Wigner molecular crystals from multielectron moiré artificial atoms}}, 
author = {Li, Hongyuan and Xiang, Ziyu and Reddy, Aidan P. and Devakul, Trithep and Sailus, Renee and Banerjee, Rounak and Taniguchi, Takashi and Watanabe, Kenji and Tongay, Sefaattin and Zettl, Alex and Fu, Liang and Crommie, Michael F. and Wang, Feng}, 
journal = {Science}, 
issn = {0036-8075}, 
doi = {10.1126/science.adk1348}, 
pages = {86--91}, 
number = {6704}, 
volume = {385}, 
}

@article{Zhang2024Engineering, 
year = {2024}, 
title = {{Engineering correlated insulators in bilayer graphene with a remote Coulomb superlattice}}, 
author = {Zhang, Zuocheng and Xie, Jingxu and Zhao, Wenyu and Qi, Ruishi and Sanborn, Collin and Wang, Shaoxin and Kahn, Salman and Watanabe, Kenji and Taniguchi, Takashi and Zettl, Alex and Crommie, Michael and Wang, Feng}, 
journal = {Nat. Mater.}, 
doi = {10.1038/s41563-023-01754-3}, 
  volume={23},
  number={2},
  pages={189--195},
}

@article{Nuckolls2024A, 
year = {2024}, 
title = {{A microscopic perspective on moiré materials}}, 
author = {Nuckolls, Kevin P. and Yazdani, Ali}, 
journal = {Nat. Rev. Mater.}, 
doi = {10.1038/s41578-024-00682-1}, 
  volume={9},
  number={7},
  pages={460--480}
}

@article{Cao2018Unconventional, 
year = {2018}, 
title = {{Unconventional superconductivity in magic-angle graphene superlattices}}, 
author = {Cao, Yuan and Fatemi, Valla and Fang, Shiang and Watanabe, Kenji and Taniguchi, Takashi and Kaxiras, Efthimios and Jarillo-Herrero, Pablo}, 
journal = {Nature}, 
issn = {0028-0836}, 
doi = {10.1038/nature26160}, 
pages = {43--50}, 
number = {7699}, 
volume = {556}
}

@article{Sharpe2019Emergent, 
year = {2019}, 
title = {{Emergent ferromagnetism near three-quarters filling in twisted bilayer graphene}}, 
author = {Sharpe, Aaron L. and Fox, Eli J. and Barnard, Arthur W. and Finney, Joe and Watanabe, Kenji and Taniguchi, Takashi and Kastner, M. A. and Goldhaber-Gordon, David}, 
journal = {Science}, 
issn = {0036-8075}, 
doi = {10.1126/science.aaw3780}, 
pages = {605--608}, 
number = {6453}, 
volume = {365}, 
}

@article{tilak2024proximity, 
  year     = {2024}, 
  title    = {Proximity induced charge density wave in a graphene/1{T}-{TaS$_2$} heterostructure}, 
  author   = {Tilak, Nikhil and Altvater, Michael and Hung, Sheng-Hsiung and Won, Choong-Jae and Li, Guohong and Kaleem, Taha and Cheong, Sang-Wook and Chung, Chung-Hou and Jeng, Horng-Tay and Andrei, Eva Y.}, 
  journal  = {Nat. Commun.}, 
  doi      = {10.1038/s41467-024-51608-y}, 
  pages    = {8056}, 
  number   = {1}, 
  volume   = {15}
}

@article{Bao2021Experimental, 
year = {2021}, 
title = {{Experimental Evidence of Chiral Symmetry Breaking in Kekulé-Ordered Graphene}}, 
author = {Bao, Changhua and Zhang, Hongyun and Zhang, Teng and Wu, Xi and Luo, Laipeng and Zhou, Shaohua and Li, Qian and Hou, Yanhui and Yao, Wei and Liu, Liwei and Yu, Pu and Li, Jia and Duan, Wenhui and Yao, Hong and Wang, Yeliang and Zhou, Shuyun}, 
journal = {Phys. Rev. Lett.}, 
doi = {10.1103/physrevlett.126.206804}, 
pages = {206804}, 
number = {20}, 
volume = {126}
}

@article{ortix2011graphene-130, 
  year     = {2011}, 
  title    = {Graphene on incommensurate substrates: Trigonal warping and emerging {Dirac} cone replicas with halved group velocity}, 
  author   = {Ortix, Carmine and Yang, Liping and Brink, Jeroen van den}, 
  journal  = {Phys. Rev. B}, 
  doi      = {10.1103/physrevb.86.081405}, 
  pages    = {081405}, 
  number   = {8}, 
  volume   = {86}
}

@article{balents2020superconductivity, 
  year     = {2020}, 
  title    = {Superconductivity and strong correlations in moiré flat bands}, 
  author   = {Balents, Leon and Dean, Cory R. and Efetov, Dmitri K. and Young, Andrea F.}, 
  journal  = {Nat. Phys.}, 
  doi      = {10.1038/s41567-020-0906-9}, 
  pages    = {725--733}, 
  number   = {7}, 
  volume   = {16}
}

@article{ponomarenko2013cloning-b61, 
  year     = {2013}, 
  title    = {Cloning of {Dirac} fermions in graphene superlattices}, 
  author   = {Ponomarenko, L. A. and Gorbachev, R. V. and Yu, G. L. and Elias, D. C. and Jalil, R. and Patel, A. A. and Mishchenko, A. and Mayorov, A. S. and Woods, C. R. and Wallbank, J. R. and Mucha-Kruczynski, M. and Piot, B. A. and Potemski, M. and Grigorieva, I. V. and Novoselov, K. S. and Guinea, F. and Fal’ko, V. I. and Geim, A. K.}, 
  journal  = {Nature}, 
  doi      = {10.1038/nature12187}, 
  pages    = {594--597}, 
  number   = {7451}, 
  volume   = {497}
}

@article{
nuckolls2025spectroscopy-726, 
  year     = {2025}, 
  title    = {Spectroscopy of the fractal {Hofstadter} energy spectrum}, 
  author   = {Nuckolls, Kevin P. and Scheer, Michael G. and Wong, Dillon and Oh, Myungchul and Lee, Ryan L. and Herzog-Arbeitman, Jonah and Watanabe, Kenji and Taniguchi, Takashi and Lian, Biao and Yazdani, Ali}, 
  journal  = {Nature}, 
  doi      = {10.1038/s41586-024-08550-2}, 
  pages    = {60--66}, 
  number   = {8053}, 
  volume   = {639}
}

@article{chen2020tunable-f63, 
  year     = {2020}, 
  title    = {Tunable correlated {Chern} insulator and ferromagnetism in a moiré superlattice}, 
  author   = {Chen, Guorui and Sharpe, Aaron L. and Fox, Eli J. and Zhang, Ya-Hui and Wang, Shaoxin and Jiang, Lili and Lyu, Bosai and Li, Hongyuan and Watanabe, Kenji and Taniguchi, Takashi and Shi, Zhiwen and Senthil, T. and Goldhaber-Gordon, David and Zhang, Yuanbo and Wang, Feng}, 
  journal  = {Nature}, 
  doi      = {10.1038/s41586-020-2049-7},  
  pages    = {56--61}, 
  number   = {7797}, 
  volume   = {579}
}

@article{woods2014commensurateincommensurate-d0c, 
  year     = {2014}, 
  title    = {Commensurate-incommensurate transition in graphene on hexagonal boron nitride}, 
  author   = {Woods, C. R. and Britnell, L. and Eckmann, A. and Ma, R. S. and Lu, J. C. and Guo, H. M. and Lin, X. and Yu, G. L. and Cao, Y. and Gorbachev, R. V. and Kretinin, A. V. and Park, J. and Ponomarenko, L. A. and Katsnelson, M. I. and Gornostyrev, Yu. N. and Watanabe, K. and Taniguchi, T. and Casiraghi, C. and Gao, H-J. and Geim, A. K. and Novoselov, K. S.}, 
  journal  = {Nat. Phys.}, 
  doi      = {10.1038/nphys2954}, 
  pages    = {451--456}, 
  number   = {6}, 
  volume   = {10}
}

@article{pasztor2019holographic-369, 
  year     = {2019}, 
  title    = {Holographic imaging of the complex charge density wave order parameter}, 
  author   = {Pasztor, Arpad and Scarfato, Alessandro and Spera, Marcello and Barreteau, Céline and Giannini, Enrico and Renner, Christoph}, 
  journal  = {Phys. Rev. Res.}, 
  doi      = {10.1103/physrevresearch.1.033114}, 
  pages    = {033114}, 
  number   = {3}, 
  volume   = {1}
}

@article{chen2020visualizing-890, 
  year     = {2020}, 
  title    = {Visualizing the Anomalous Charge Density Wave States in Graphene/{NbSe$_2$} Heterostructures}, 
  author   = {Chen, Yu and Wu, Lishu and Xu, Hai and Cong, Chunxiao and Li, Si and Feng, Shun and Zhang, Hongbo and Zou, Chenji and Shang, Jingzhi and Yang, Shengyuan A. and Loh, Kian Ping and Huang, Wei and Yu, Ting}, 
  journal  = {Adv. Mater.}, 
  doi      = {10.1002/adma.202003746}, 
  pages    = {2003746}, 
  number   = {45}, 
  volume   = {32}
}

@article{bischoff2017nanoscale-01b, 
  year     = {2017}, 
  title    = {Nanoscale Phase Engineering of Niobium Diselenide}, 
  author   = {Bischoff, Felix and Auwärter, Willi and Barth, Johannes V and Schiffrin, Agustin and Fuhrer, Michael and Weber, Bent}, 
  journal  = {Chem. of Mater.}, 
  doi      = {10.1021/acs.chemmater.7b03061}, 
  pages    = {9907--9914}, 
  number   = {23}, 
  volume   = {29}
}

@article{slater1954simplified-305, 
  year     = {1954}, 
  title    = {Simplified {LCAO} Method for the Periodic Potential Problem}, 
  author   = {Slater, J. C. and Koster, G. F.}, 
  journal  = {Phys. Rev.}, 
  doi      = {10.1103/physrev.94.1498}, 
  pages    = {1498--1524}, 
  number   = {6}, 
  volume   = {94}
}

@article{kim2024effect-e04, 
  year     = {2024}, 
  title    = {Effect of Charge Density Wave on the Electronic Transport in Graphene}, 
  author   = {Kim, Boram and Li, Jinshu and Park, Jeehoon and Lim, Hongsik and Myeong, Gyuho and Shin, Wongil and Kim, Seungho and Jin, Taehyeok and Zhang, Qi and Sung, Kyunghwan and Le, Duc Duy and Watanabe, Kenji and Taniguchi, Takashi and Yang, Chan-Ho and Hwang, Euyheon and Cho, Sungjae}, 
  journal  = {ACS Appl. Electron. Mater.}, 
  doi      = {10.1021/acsaelm.3c01560}, 
  pages    = {1174--1180}, 
  number   = {2}, 
  volume   = {6}
}

@article{zhao2022moiré-82f, 
  year     = {2022}, 
  title    = {Moiré enhanced charge density wave state in twisted 1{T}-{TiTe$_2$}/1{T}-{TiSe$_2$} heterostructures}, 
  author   = {Zhao, Wei-Min and Zhu, Li and Nie, Zhengwei and Li, Qi-Yuan and Wang, Qi-Wei and Dou, Li-Guo and Hu, Ju-Gang and Xian, Lede and Meng, Sheng and Li, Shao-Chun}, 
  journal  = {Nat. Mater.}, 
  doi      = {10.1038/s41563-021-01167-0}, 
  pages    = {284--289}, 
  number   = {3}, 
  volume   = {21}
}

@article{yoo2019atomic-562, 
  year     = {2019}, 
  title    = {Atomic and electronic reconstruction at the van der {Waals} interface in twisted bilayer graphene}, 
  author   = {Yoo, Hyobin and Engelke, Rebecca and Carr, Stephen and Fang, Shiang and Zhang, Kuan and Cazeaux, Paul and Sung, Suk Hyun and Hovden, Robert and Tsen, Adam W. and Taniguchi, Takashi and Watanabe, Kenji and Yi, Gyu-Chul and Kim, Miyoung and Luskin, Mitchell and Tadmor, Ellad B. and Kaxiras, Efthimios and Kim, Philip}, 
  journal  = {Nat. Mater.}, 
  doi      = {10.1038/s41563-019-0346-z}, 
  pages    = {448--453}, 
  number   = {5}, 
  volume   = {18}
}

@article{liu2021monolayer-ebe, 
  year     = {2021}, 
  title    = {Monolayer 1{T}-{NbSe$_2$} as a 2{D}-correlated magnetic insulator}, 
  author   = {Liu, Mengke and Leveillee, Joshua and Lu, Shuangzan and Yu, Jia and Kim, Hyunsue and Tian, Cheng and Shi, Youguo and Lai, Keji and Zhang, Chendong and Giustino, Feliciano and Shih, Chih-Kang}, 
  journal  = {Sci. Adv.}, 
  doi      = {10.1126/sciadv.abi6339}, 
  pages    = {eabi6339}, 
  number   = {47}, 
  volume   = {7}
}

@article{nakata2021robust-bae, 
  year     = {2021}, 
  title    = {Robust charge-density wave strengthened by electron correlations in monolayer {1T}-{TaSe$_2$} and {1T}-{NbSe$_2$}}, 
  author   = {Nakata, Yuki and Sugawara, Katsuaki and Chainani, Ashish and Oka, Hirofumi and Bao, Changhua and Zhou, Shaohua and Chuang, Pei-Yu and Cheng, Cheng-Maw and Kawakami, Tappei and Saruta, Yasuaki and Fukumura, Tomoteru and Zhou, Shuyun and Takahashi, Takashi and Sato, Takafumi}, 
  journal  = {Nat. Commun.}, 
  doi      = {10.1038/s41467-021-26105-1}, 
  pages    = {5873}, 
  number   = {1}, 
  volume   = {12}
}

@article{wallbank2013generic-397, 
  year     = {2013}, 
  title    = {Generic miniband structure of graphene on a hexagonal substrate}, 
  author   = {Wallbank, J. R. and Patel, A. A. and Mucha-Kruczyński, M. and Geim, A. K. and Fal'ko, V. I.}, 
  journal  = {Phys. Rev. B}, 
  doi      = {10.1103/physrevb.87.245408}, 
  pages    = {245408}, 
  number   = {24}, 
  volume   = {87}
}

@article{qiu2025giant-d55, 
  year     = {2025}, 
  title    = {{Giant Splitting of Folded Dirac Bands in Kekulé-ordered Graphene with Eu Intercalation}}, 
  author   = {Qiu, Xiaodong and Zhu, Tongshuai and Fan, Zhenjie and Wang, Kaili and Mu, Yuyang and Yang, Bin and Wu, Di and Zhang, Haijun and Wang, Can and Wang, Huaiqiang and Zhang, Yi}, 
  journal  = {arXiv}, 
  doi      = {10.48550/arxiv.2509.05633}, 
}

@article{Du2024, 
  year     = {2024}, 
title      ={Nonlinear physics of moiré superlattices},
  author   = {Du, Luojun Du and Huang, Zhiheng and 
Zhang, Jin and Ye, Fangwei and Dai, Qing and Deng, Hui and 
Zhang , Guangyu  and Sun, Zhipei}, 
  journal  = {Nat. Mater.}, 
  doi      = {10.1038/s41563-024-01951-8},
  pages    = {1179}, 
  volume   = {23}
}

@article{lu2024fractional-60e, 
  year     = {2024}, 
  title    = {Fractional quantum anomalous {Hall} effect in multilayer graphene}, 
  author   = {Lu, Zhengguang and Han, Tonghang and Yao, Yuxuan and Reddy, Aidan P. and Yang, Jixiang and Seo, Junseok and Watanabe, Kenji and Taniguchi, Takashi and Fu, Liang and Ju, Long}, 
  journal  = {Nature}, 
  doi      = {10.1038/s41586-023-07010-7}, 
  pages    = {759--764}, 
  number   = {8000}, 
  volume   = {626}
}

@article{tanaka2025superfluid-65f, 
  year     = {2025}, 
  title    = {Superfluid stiffness of magic-angle twisted bilayer graphene}, 
  author   = {Tanaka, Miuko and Wang, Joel I-j. and Dinh, Thao H. and Rodan-Legrain, Daniel and Zaman, Sameia and Hays, Max and Almanakly, Aziza and Kannan, Bharath and Kim, David K. and Niedzielski, Bethany M. and Serniak, Kyle and Schwartz, Mollie E. and Watanabe, Kenji and Taniguchi, Takashi and Orlando, Terry P. and Gustavsson, Simon and Grover, Jeffrey A. and Jarillo-Herrero, Pablo and Oliver, William D.}, 
  journal  = {Nature}, 
  doi      = {10.1038/s41586-024-08494-7}, 
  pages    = {99--105}, 
  number   = {8049}, 
  volume   = {638}
}

@article{barré2024engineering-b09, 
  year     = {2024}, 
  title    = {Engineering interlayer hybridization in van der {Waals} bilayers}, 
  author   = {Barre, Elyse and Dandu, Medha and Kundu, Sudipta and Sood, Aditya and Jornada, Felipe H. da and Raja, Archana}, 
  journal  = {Nat. Rev. Mater.}, 
  doi      = {10.1038/s41578-024-00666-1}, 
  pages    = {499--508}, 
  number   = {7}, 
  volume   = {9}
}

@article{huang2025doped-cf4, 
  year     = {2025}, 
  title    = {{Doped Mott Phase and Charge Correlations in Monolayer 1T-NbSe$_2$}}, 
  author   = {Huang, Xin and Lado, Jose L. and Sainio, Jani and Liljeroth, Peter and Ganguli, Somesh Chandra}, 
  journal  = {Phys. Rev. Lett.}, 
  doi      = {10.1103/physrevlett.134.046504}, 
  pages    = {046504}, 
  number   = {4}, 
  volume   = {134}
}

@article{tian2024dominant-b2e, 
  year     = {2024}, 
  title    = {Dominant 1/3-filling correlated insulator states and orbital geometric frustration in twisted bilayer graphene}, 
  author   = {Tian, Haidong and Codecido, Emilio and Mao, Dan and Zhang, Kevin and Che, Shi and Watanabe, Kenji and Taniguchi, Takashi and Smirnov, Dmitry and Kim, Eun-Ah and Bockrath, Marc and Lau, Chun Ning}, 
  journal  = {Nat. Phys.}, 
  doi      = {10.1038/s41567-024-02546-5}, 
  pages    = {1407--1412}, 
  number   = {9}, 
  volume   = {20}
}

@article{su2025moirédriven-896, 
  year     = {2025}, 
  title    = {Moiré-driven topological electronic crystals in twisted graphene}, 
  author   = {Su, Ruiheng and Waters, Dacen and Zhou, Boran and Watanabe, Kenji and Taniguchi, Takashi and Zhang, Ya-Hui and Yankowitz, Matthew and Folk, Joshua}, 
  journal  = {Nature}, 
  doi      = {10.1038/s41586-024-08239-6}, 
  pages    = {1084--1089}, 
  number   = {8048}, 
  volume   = {637}
}

@article{zhao2024emergence-34c, 
  year     = {2024}, 
  title    = {Emergence of ferromagnetism at the onset of moiré Kondo breakdown}, 
  author   = {Zhao, Wenjin and Shen, Bowen and Tao, Zui and Kim, Sunghoon and Knüppel, Patrick and Han, Zhongdong and Zhang, Yichi and Watanabe, Kenji and Taniguchi, Takashi and Chowdhury, Debanjan and Shan, Jie and Mak, Kin Fai}, 
  journal  = {Nat. Phys.}, 
  doi      = {10.1038/s41567-024-02636-4}, 
  pages    = {1772--1777}, 
  number   = {11}, 
  volume   = {20}
}

@article{checkelsky2024flat-c0b, 
  year     = {2024}, 
  title    = {Flat bands, strange metals and the {Kondo} effect}, 
  author   = {Checkelsky, Joseph G. and Bernevig, B. Andrei and Coleman, Piers and Si, Qimiao and Paschen, Silke}, 
  journal  = {Nat. Rev. Mater.}, 
  doi      = {10.1038/s41578-023-00644-z}, 
  volume   = {9},
  pages    = {509--526}
}

@article{lai2025moire-16a, 
  year     = {2025}, 
  title    = {Moiré periodic and quasiperiodic crystals in heterostructures of twisted bilayer graphene on hexagonal boron nitride}, 
  author   = {Lai, Xinyuan and Li, Guohong and Coe, Angela M. and Pixley, Jedediah H. and Watanabe, Kenji and Taniguchi, Takashi and Andrei, Eva Y.}, 
  journal  = {Nat. Mater.}, 
  doi      = {10.1038/s41563-025-02222-w}, 
  pages    = {1019--1026}, 
  number   = {7}, 
  volume   = {24}
}

@article{zhang2025moiré-128, 
  year     = {2025}, 
  title    = {Moiré enhanced flat band in rhombohedral graphene}, 
  author   = {Zhang, Hongyun and Lu, Jinxi and Liu, Kai and Wang, Yijie and Wang, Fei and Wu, Size and Chen, Wanying and Cai, Xuanxi and Watanabe, Kenji and Taniguchi, Takashi and Avila, Jose and Dudin, Pavel and Watson, Matthew D. and Louat, Alex and Sato, Takafumi and Yu, Pu and Duan, Wenhui and Song, Zhida and Chen, Guorui and Zhou, Shuyun}, 
  journal  = {Nat. Mater.}, 
  doi      = {10.1038/s41563-025-02416-2}, 
  pages    = {1--7}
}

@article{oudich2024engineered-e09, 
  year     = {2024}, 
  title    = {Engineered moiré photonic and phononic superlattices}, 
  author   = {Oudich, Mourad and Kong, Xianghong and Zhang, Tan and Qiu, Chengwei and Jing, Yun}, 
  journal  = {Nat. Mater.}, 
  doi      = {10.1038/s41563-024-01950-9}, 
  pages    = {1169--1178}, 
  number   = {9}, 
  volume   = {23}
}

@article{guo2025superconductivity-263, 
  year     = {2025}, 
  title    = {Superconductivity in 5.0° twisted bilayer {WSe$_2$}}, 
  author   = {Guo, Yinjie and Pack, Jordan and Swann, Joshua and Holtzman, Luke and Cothrine, Matthew and Watanabe, Kenji and Taniguchi, Takashi and Mandrus, David G. and Barmak, Katayun and Hone, James and Millis, Andrew J. and Pasupathy, Abhay and Dean, Cory R.}, 
  journal  = {Nature}, 
  doi      = {10.1038/s41586-024-08381-1}, 
  pages    = {839--845}, 
  number   = {8047}, 
  volume   = {637}
}

@article{torre2025advanced-b8f, 
  year     = {2025}, 
  title    = {{Advanced Characterization of the Spatial Variation of Moiré Heterostructures and Moiré Excitons}}, 
  author   = {Torre, Alberto de la and Kennes, Dante M. and Malic, Ermin and Kar, Swastik}, 
  journal  = {Small}, 
  doi      = {10.1002/smll.202401474}, 
  pages    = {2401474}, 
  number   = {28}, 
  volume   = {21}
}

@article{wood1964vocabulary-bfd, 
  year     = {1964}, 
  title    = {{Vocabulary of Surface Crystallography}}, 
  author   = {Wood, Elizabeth A}, 
  journal  = {J. Appl. Phys.}, 
  doi      = {10.1063/1.1713610}, 
  pages    = {1306--1312}, 
  number   = {4}, 
  volume   = {35}
}
\clearpage

\makeatletter
\renewcommand\section{\@startsection {section}{1}{\z@}%
  {-3.5ex \@plus -1ex \@minus -.2ex}%
  {2.3ex \@plus .2ex}%
  {\normalfont\normalsize\bfseries}} % Remove \uppercase
\makeatother

%===============================================================================
\section*{Acknowledgments}
%===============================================================================
We would like to thank Tim Wehling, Tobias Wichmann and F.~Stefan Tautz for their fruitful discussions and valuable feedback. We also gratefully acknowledge computing time granted through JARA on the supercomputer JURECA and the JURECA DC Evaluation platform at Forschungszentrum Jülich. 
Furthermore, the authors are grateful to the Helmholtz Nano Facility for its support regarding sample fabrication. 
K.~J., F.~L., D.~M.~K., J.~M.-C.,  and M.~T.\ acknowledge funding by the Deutsche Forschungsgemeinschaft (DFG, German Research Foundation) within the SPP 2244 (Projects 443416235, 422707584 and 443274199) and by a Major Instrumentation fund (INST 222/1296-1 FUGG). 
J.~M.-C., and F.~L.\ acknowledge funding from the Bavarian Ministry of Economic Affairs, Regional Development, and Energy within Bavaria's High-Tech Agenda Project “Bausteine für das Quantencomputing auf Basis topologischer Materialien mit experimentellen und theoretischen Ansätzen”.
J.~M.-C.\ acknowledges funding from the Alexander von Humboldt Foundation. L.~K.\ acknowledges support through the Würzburg-Dresden Cluster of Excellence on Complexity and Topology in Quantum Matter – ctd.qmat, Project-ID 390858490 – EXC 2147, and through the Research Unit QUAST, Project-ID 449872909 – FOR5249.  M.~T.\ acknowledges funding from the Heisenberg Program (TE 833/2) of the German Research Foundation. F.~L.\ acknowledges financial support from Germany's Excellence Strategy-Cluster of Excellence Matter and Light for Quantum Computing (ML4Q) through an Independence Grant. Z.\ A.\ H.\ G acknowledges support through the Glasstone Research Fellowship in Materials, University of Oxford.

\subsection*{Author contributions}
K.~J.\, J.~M.-C, and M.~T.\ conceived the project. K.~J. fabricated the samples. K.~J., J.~Z.\ and J.~M.-C.\ performed the STM experiments. K.~J. and  M.~T.\ analyzed the data. L.~K., Z.~A.~H.~G., and D.~M.~K.\ performed the DFT and the geometry model calculations. F.~L., D.~M,~K., J.~M.-C., and M.~T. supervised the project. K.~J., and M.~T.\ wrote the manuscript. All authors discussed and commented on the manuscript.

\subsection*{Competing financial interests}
The authors declare no competing financial interests.

%\bibliography{Refs.bib}
\end{document}